

\input harvmac
\noblackbox
\pageno=0\nopagenumbers\tolerance=10000\hfuzz=5pt
\line{\hfill CERN-TH/95-1}
\vskip 24pt
\centerline{\bf SUMMATION OF LEADING LOGARITHMS AT SMALL $x$}
\vskip 36pt\centerline{Richard~D.~Ball\footnote{*}{On leave
 from a Royal Society University Research Fellowship.}
 and Stefano~Forte\footnote{\dag}{On leave
 from INFN, Sezione di Torino, Italy.}}
\vskip 12pt
\centerline{\it Theory Division, CERN,}
\centerline{\it CH-1211 Gen\`eve 23, Switzerland.}
\vskip 36pt
{\centerline{\bf Abstract}
\medskip\narrower
\ninepoint\baselineskip=9pt plus 2pt minus 1pt
\lineskiplimit=1pt \lineskip=2pt
We show how perturbation theory may be reorganized to give splitting
functions which include order by order convergent sums of all leading
logarithms of $x$. This gives a leading twist evolution equation for
parton distributions which sums all leading logarithms of $x$ and
$Q^2$, allowing stable perturbative evolution down to arbitrarily
small values of $x$. Perturbative evolution then generates the double
scaling rise of $F_2$ observed at HERA, while in the formal limit
$x\to 0$ at fixed $Q^2$ the Lipatov $x^{-\lambda}$ behaviour is
eventually reproduced. We are thus able to explain why leading order
perturbation theory works so well in the HERA region.
\smallskip}
\vskip 20pt \centerline{Submitted to: {\it Physics Letters B}}
\vskip 24pt
\line{CERN-TH/95-1\hfill}
\line{January 1995\hfill}

\vfill\eject \footline={\hss\tenrm\folio\hss}


\def\Re{\,\hbox{Re}\,}

\def\rhs{right hand side}

\def\neath#1#2{\mathrel{\mathop{#1}\limits{#2}}}
\def\frac#1#2{{{#1}\over {#2}}}
\def\half{\hbox{${1\over 2}$}}

\def\smallfrac#1#2{\hbox{${{#1}\over {#2}}$}}

\def\MeV{{\rm MeV}}\def\GeV{{\rm GeV}}

\catcode`@=11 
\def\slash#1{\mathord{\mathpalette\c@ncel#1}}
 \def\c@ncel#1#2{\ooalign{$\hfil#1\mkern1mu/\hfil$\crcr$#1#2$}}
\def\lsim{\mathrel{\mathpalette\@versim<}}
\def\gsim{\mathrel{\mathpalette\@versim>}}
 \def\@versim#1#2{\lower0.2ex\vbox{\baselineskip\z@skip\lineskip\z@skip
       \lineskiplimit\z@\ialign{$\m@th#1\hfil##$\crcr#2\crcr\sim\crcr}}}
\catcode`@=12 

\def\PR{{\it Phys.~Rev.~}}

\def\NP{{\it Nucl.~Phys.~}}
\def\NPBPS{{\it Nucl.~Phys.~B (Proc.~Suppl.)~}}
\def\PL{{\it Phys.~Lett.~}}
\def\PRep{{\it Phys.~Rep.~}}
\def\AP{{\it Ann.~Phys.~}}

\def\SJNP{{\it Sov.~Jour.~Nucl.~Phys.~}}

\def\vol#1{{\bf #1}}\def\vyp#1#2#3{\vol{#1} (#2) #3}


The evolution of structure functions at small $x$ is
expected to be problematic within the conventional framework \ref\GPGW
{H.~Georgi and H.D.~Politzer,
        \PR\vyp{D9}{1974}{416}\semi D.~Gross and F.~Wilczek,
        \PR\vyp{D9}{1974}{980}.}
of perturbative QCD, in that the usual perturbative
evolution appears to be unstable at small $x$. This expectation rests on
the observation that singlet anomalous dimensions grow in an
unbounded way as $N\to 0$ \ref\Gross
{D.~Gross in the proceedings of the XVII Intern. Conf. on
High Energy Physics, London, 1974
                       (published by SRC, Rutherford Lab.) and
lectures given at Les Houches, Session XXVIII, 1975, published in
``Methods in Field Theory'', ed. R.~Balian and J.~Zinn-Justin
(North-Holland, 1976).}
(specifically, at order $m$ in perturbation theory
$\gamma^{gg}_N(\alpha)\sim \big(\frac{\alpha}{N}\big)^m$). Indeed, the leading
order term drives a corresponding growth in the singlet structure
function as both $\frac{1}{x}$ and $Q^2$ increase \ref\DGPTWZ{A.~De~Rujula,
S.L.~Glashow, H.D.~Politzer, S.B.~Treiman, F.~Wilczek and A.~Zee,
\PR\vyp{D10}{1974}{1649}.}. However if $\frac{1}{x}$ increases much
more rapidly than $t\equiv \ln Q^2/\Lambda^2$, higher order terms
become increasingly important, and it is no longer clear whether
perturbation theory can be trusted\Gross. Hence it is often claimed
that the usual evolution equations \GPGW\ must then be abandoned in favour of
equations \ref\Lip{L.N.~Lipatov, \SJNP\vyp{23}{1976}{338}\semi
          V.S.~Fadin, E.A.~Kuraev and L.N.~Lipatov,
       \PL\vyp{60B}{1975}{50};
       {\it Sov. Phys. JETP~}\vyp{44}{1976}{443};\vyp{45}{1977}{199}\semi
          Y.Y.~Balitski and L.N.Lipatov, \SJNP\vyp{28}{1978}{822}.}
which resum the logs of $\frac{1}{x}$ at fixed $t$, resulting in an
even stronger (power-like) growth.

\nref\ZEUSc{ZEUS Collab., M.~Lancaster, talk at the 27th
International Conference on High Energy Physics, July 1994, Glasgow,
and DESY-94-143}
\nref\Honec{H1 Collab., talk at the 27th
International Conference on High Energy Physics, July 1994, Glasgow.}
Strong scaling violations at large $Q^2$ accompanied by a
corresponding increase at small values of $x$ are indeed observed in
recent measurements of the proton structure at
HERA\refs{\ZEUSc,\Honec}. The leading order perturbative prediction \DGPTWZ\
can thus now be tested empirically: the structure
function data should display a double scaling behavior\ref\DAS{R.~D.~Ball and
S.~Forte, \PL\vyp{B335}{1994}{77}.} in the two variables
$\sigma\sim\sqrt{\ln\smallfrac{1}{x}\ln t}$ and
$\rho\sim\sqrt{\ln\smallfrac{1}{x}\big/\ln t}$, rising
linearly in the former while remaining relatively
independent of the latter. This scaling follows directly from the fact
that the leading order small-$x$ evolution equations reduce asymptotically
to an isotropic wave equation in the plane of $\ln\frac{1}{x}$ and
$\ln t$. In fact not only is the scaling observed\DAS, but even
the slope of the rise turns out to be remarkably close to the leading
order prediction \ref\Test
{R.~D.~Ball and S.~Forte, \PL\vyp{B336}{1994}{77}.}.
These results remain stable upon the inclusion of two loop corrections
\ref\Mont{R.D.~Ball and S.~Forte, CERN-TH.7422/94, {\tt hep-ph/9409374},
        to be published in the proceedings of {\it ``QCD94''},
        Montpellier, July 1994 (\NPBPS).},
and it now seems possible to detect two loop effects in the
data through a small reduction in the slope.

\nref\Blois{R.D.~Ball and S.~Forte,
            CERN-TH.7421/94, {\tt hep-ph/9409373},
        to be published in {\it ``The Heart of the Matter''},
        VIth Rencontre de Blois, June 1994 (Editions Fronti\`eres).}
\nref\BFKLad{T.~Jaroszewicz, \PL\vyp{B116}{1982}{291}.}
\nref\BFKLadproof{S.~Catani, F.~Fiorani and G.~Marchesini,
\PL\vyp{B234}{1990}{339}; \NP\vyp{B336}{1990}{18}\semi
  S.~Catani, F.~Fiorani, G.~Marchesini and G.~Oriani,
\NP\vyp{B361}{1991}{645}.}
\nref\CCH{S.~Catani, M.~Ciafaloni and F.~Hautmann,
            \PL\vyp{B242}{1990}{97}; \NP\vyp{B366}{1991}{135};
            \PL\vyp{B307}{1993}{147}.}
\nref\CHad{ S.~Catani \& F.~Hautmann, \PL\vyp{B315}{1993}{157},
                  preprint Cavendish-HEP-94-01.\semi
        F.~Hautmann, to be published in the proceedings of {\it ``QCD94''},
        Montpellier, July 1994 (\NPBPS).}
\nref\Catrev{ S.~Catani, DFF~207/6/94, talk given at {\it Les
Rencontres de Physique de La Vall\'ee d'Aoste}, La Thuile, 1994.}

All this suggests that it should be possible to keep the
behaviour of the perturbative expansion under control by summing
leading logs of $\frac{1}{x}$ as well as logs of $Q^2$, while
remaining within the standard framework of perturbative evolution \Blois.
The double asymptotic scaling behaviour observed in
the HERA data should then emerge as the dominant behaviour in some
well determined region of the $x$--$t$ plane.
Such a programme is possible since the leading
singularities in both $\gamma^{gg}_N$ and $\gamma_N^{qg}$ have now
been explicitly computed to all orders in $\alpha_s/N$
\refs{\BFKLad-\Catrev}. However in a previous study\ref\EKL{R.~K.~Ellis,
Z.~Kunszt and E.~M.~Levin, \NP\vyp{B420}{1994}{517}.}, in which
three and four loop singularities were retained, it appeared that
instabilities may develop which signal the breakdown of perturbation
theory. Here we will explain in detail
how the perturbative expansion of anomalous dimensions can be
reorganized at small-$x$ in a way which is consistent with the
renormalization group, but in which the leading logs of $\frac{1}{x}$
are summed to all orders. We will then prove explicitly that this leads to
stable perturbative evolution equations for parton
distributions at all positive values of $x$, in which all corrections
are either of higher order in $\alpha_s$ or are higher twist, and
which thus breaks down only at low values of $Q^2$. Solution
of these equations will then enable us to determine in precisely which
region of the $x$--$t$ plane double scaling should remain valid, and
whether it will be possible to observe the stronger power-like rise
characteristic of solutions of the BFKL equation.

According to the operator product expansion and renormalization group,
the Mellin moments $f_N^i(t)\equiv\int_0^1 dx x^{N} f^i(x;t)$ of the
parton distribution functions $f^i(x;t)$ evolve multiplicatively with
$t$:
\eqn\evol{\frac{d}{dt}f_N^i(t)=\sum_j\gamma^{ij}_N(\alpha_s(t))f_N^j(t),}
where $\gamma^{ij}_N(\alpha)$ are the anomalous dimensions of local
operators \GPGW. The perturbative evolution then sums
all logarithms of $Q^2$, so the $O(\alpha)$ and
$O(\alpha^2)$ terms are known as the leading and subleading log
approximations respectively. Perturbative expansion of the anomalous
dimensions is justified by factorization theorems; these have been
shown recently to apply to all orders even at small $x$
\refs{\CCH-\Catrev}. The inverse Mellin
transform of \evol\ gives
\eqn\evolAP{\frac{d}{dt}f^i(x;t)=\sum_j\int_x^1 dy P^{ij}(y;t)
f^j(\smallfrac{x}{y};t),}
where the splitting functions $P^{ij}(x;t)$ are related to the anomalous
dimensions by\foot{This definition differs from the
usual one by a factor of $\smallfrac{\alpha_s}{2\pi}$; also our moment
variable $N$ differs by one from the usual one so that for us the
first moment is that with $N=0$.}
\eqn\mel
{\gamma_N^{ij}(\alpha_s(t))=\int_0^1\!\!dx\,\, x^{N} P^{ij}(x;t).}
This version of the evolution equation is more physical because it
only involves integration over the physically accessible region $y>x$ \ref\AP
{G.~Parisi, \PL\vyp{50B}{1974}{367}\semi
G.~Parisi, {\it Proc. 11th Rencontre de Moriond,}
ed. J.~Tran~Thanh~Van, ed. Fronti\`eres, 1976\semi
G.~Altarelli and G.~Parisi, \NP\vyp{B126}{1977}{298}\semi
G.~Altarelli, \PRep\vyp{81}{1981}{1}.}.

\nfig\figexp{The terms summed in
the various expansions of the anomalous dimensions and associated
splitting functions: a) the standard (large-$x$) expansion \laxexpad, b)
the small-$x$ expansion \smxexpad, and c) the `double-leading'
expansion \dlexpad. Leading, sub-leading and sub-sub-leading terms are
indicated by the solid, dashed and dotted lines respectively; $m$
denotes the order in $\alpha$, while $n$ denotes the order in $1/N$.
Singular terms are marked as crosses, while
terms whose coefficients known at present (for $\gamma^{gg}_N$)
are marked by circles: the term which leads to double scaling is
marked with a star.}
At small $x$
it is of course necessary to take into account the fact that, if
$1\over x$ is large enough, Feynman diagrams which contain powers
of $\ln{1\over x}$ may be just as important as those with powers of
$\ln Q^2$. It is thus no longer appropriate to organize the
perturbative expansion of the splitting functions in powers of
$\alpha_s$ only. In the Mellin space equation \evol\ these extra
logarithms are due to the presence of poles in the anomalous
dimensions as $N\to 0$. Now, while nonsinglet
anomalous dimensions are regular as $N\to 0$, singlet anomalous
dimensions computed at leading order in $\alpha$ have a ${1\over N}$
singularity; at $m-th$ order they have at most (at least in reasonable
renormalization schemes) an $N^{-m}$ singularity \refs{\BFKLad-\Catrev}.
Hence, separating out these singularities, and suppressing the flavor
indices $i,j$ for clarity, the anomalous dimensions
admit the perturbative expansion (see \figexp a)
\eqn\laxexpad
{\gamma_N(\alpha)= \sum_{m=1}^\infty \alpha^m
\sum_{n=-\infty}^m A^m_n N^{-n}
=\sum_{m=1}^\infty \alpha^m
\bigg(\sum_{n=1}^m A^m_n N^{-n} + \bar\gamma_N^{(m)}\bigg),}
where $A^m_n$ are simply numerical coefficients, and
$\bar\gamma_N^{(m)}$ are all regular as $N\to 0$. This
then corresponds to the expansion
\eqn\laxexpsf
{P(x;t)=\sum_{m=1}^\infty \big(\alpha_s(t)\big)^m
\bigg(\frac{1}{x}\sum_{n=1}^m A^m_n\frac{\ln^{n-1}\smallfrac{1}{x}}{(n-1)!}
+ \bar P^{(m)}(x)\bigg)}
of the splitting function, where $\bar P^{(m)}(x)$ are regular as
$x\to 0$. At small $x$ the usefulness of this expansion in powers of
$\alpha$ is spoilt by the logs of $1/x$ which can compensate for the
smallness of $\alpha_s(t)$.

However this is not the only way to order the expansion: if
instead of expanding in $\alpha$ and then in $N$, as in
\laxexpad, we choose instead to expand in $\alpha$ and $\alpha/N$ (see
\figexp b), the anomalous dimensions take the form
\eqn\smxexpad
{\gamma_N(\alpha)= \sum_{m=1}^\infty \alpha^{m-1}
\Big(\sum_{n=2-m}^\infty A^{n+m-1}_n \bigg(\frac{\alpha}{N}\bigg)^n\Big) ,}
which corresponds to the splitting function
\eqn\smxexpsf
{\eqalign{P(x;t)&=\sum_{m=1}^\infty \big(\alpha_s(t)\big)^{m-1}
\bigg(\frac{1}{x}\sum_{n=1}^\infty A^{n+m-1}_n \frac{\big(\alpha_s(t)\big)^n
\ln^{n-1}\smallfrac{1}{x}}{(n-1)!}\cr
&\qquad+ \sum_{q=0}^{m-2} A^{m-q-1}_{-q} \big(\alpha_s(t)\big)^{-q}
\frac{d^q}{dx^q}\delta(1-x) \bigg).\cr}}
While not so useful at large $x$, at small $x$ this is clearly the
most appropriate expansion, since at each order in $\alpha_s$ all the
leading logs of $1/x$ have been summed up.

Solving the evolution equations with the usual expansion
eq.~\laxexpsf\ of the splitting functions corresponds at leading order
($m=1$) to summing up all logs of the form
\eqn\lls
{\alpha_s^p (\ln Q^2)^q\left(\ln{1\over x}\right)^r }
with $q=p$, and $0\le r\le p$. If instead only the sum
of leading singularities eq.~\smxexpsf\ (again with $m=1$) is included
in the splitting functions, then solving the evolution equations sums
the leading logs \lls\ with $r=p$, and $1\le
q\le p$. Of course, if contributions which are higher order in
$\alpha_s$ are included in either expansion of the splitting function,
eventually all leading logs of both $1\over x$ and $Q^2$
are included. However while in the former logs of $Q^2$ are considered
leading, in the latter logs of $1\over x$ are leading, the roles of
$q$ and $r$ in \lls\ being interchanged. Both expansions are
consistent with the renormalization group in the sense that a
change in the scale at which the $m$-th order contribution to the
expansion is evaluated is equivalent to a change in the $m+1$-th order term.

It is now easy to see how to construct an intermediate ``double leading''
expansion which includes all leading logs, i.e. one such that
perturbative evolution sums all terms with $1\le q\le p$,
$0\le r\le p$, and $1\le p\le q+r$, so that each extra power of
$\alpha_s$ is accompanied by a log of either $1\over x$ or $Q^2$ or both.
To do this, the anomalous dimensions are expanded as (see \figexp c)
\eqn\dlexpad
{\gamma_N(\alpha)= \sum_{m=1}^\infty \alpha^{m-1}
\bigg(\sum_{n=1}^\infty A^{n+m-1}_n \bigg(\frac{\alpha}{N}\bigg)^n
+ \alpha \bar\gamma_N^{(m)} \bigg).}
Each subsequent order of the expansion contains an extra power
of $\alpha$ in comparison to the previous one, so the scheme is still
consistent with the renormalization group.
Of course, any number of  renormalization group consistent expansions which
interpolate between the double leading one (eq.~\dlexpad)
and the large $x$ (eq.~\laxexpad) and small $x$ (eq.~\smxexpad) ones can also
be constructed.
The important point is that if we choose an expansion which is appropriate
at small $x$ (say, the extreme small $x$ one \smxexpad), then each subsequent
order of the expansion is genuinely of order $\alpha$ as compared to the
previous one, all logarithms having been included in the
coefficients. When the scale increases, the higher order contributions
are then asymptotically small.

The double scaling behaviour described in refs.~\refs{\DAS,\Test},
and observed at HERA\refs{\ZEUSc,\Honec}, is governed by the
pivotal $m=n=1$ term in the various expansions
\laxexpad,\smxexpad\ and \dlexpad, with $(A_1^1)^{gg}=\gamma^2\beta_0/4\pi$
determining the slope of the rise of $F_2^p$ at small $x$ and large
$Q^2$. The subleading $A_0^1$ terms determine $\delta$ (as defined
in ref.~\DAS) and the
relative normalization of the quark and gluon distributions, while
$A_2^2=0$: the leading two-loop term $A_1^2$ is responsible for a
small but observable reduction in the steepness of the slope \Mont.
Thus at the intermediate range of $x$ and $t$ currently being explored at
HERA the double leading expansion \dlexpad\ is the most appropriate,
and should be used for a detailed comparison
to the data. Here we will instead be interested in
the small $x$ limit and will thus henceforth concentrate on the small
$x$ expansions eqns.~\smxexpad\ and \smxexpsf, which are sufficient to
demonstrate the novel features of perturbative evolution in this
kinematic regime.

Before proceeding further, we must confront the convergence issue
with which we began the paper. Consider first the leading term in the
series \smxexpad: since both $\gamma_N^{qg}$
and $\gamma_N^{qq}$ vanish at leading order, and\foot{Following the usual
conventions, $C_A = 3$, $C_F = \smallfrac{4}{3}$, $T_R = \half$ for QCD with
three colors.}
$\gamma_N^{gq}=(C_F/C_A)\gamma_N^{gg}$ \CHad, we consider explicitly
\eqn\ggls
{\gamma_N^{gg}(\alpha)=\sum_{n=1}^\infty (A_n^n)^{gg} \left(
{\alpha\over N}\right)^n + O(\alpha)
\equiv (4\ln 2)\inv A(\lambda/N) + O(\alpha),}
where
\eqn\lipex
{\lambda(\alpha)=4\ln 2 \bigg(\frac{C_A\alpha}{\pi}\bigg).}
Now, it is well known (see e.g. ref.~\Catrev) that this
series develops a branch-point singularity at $N=\lambda$ \Lip, so
that while the series
converges for $|N|>\lambda$, it is not even Borel summable
if $\Re N<\lambda$ . This seems to pose an insurmountable problem
for the perturbative approach to small $x$ evolution: the
series which defines the leading coefficient in expansion \smxexpad\
of the anomalous dimension, which was supposed to be useful for small
$N$, is actually only well defined when $N$ is sufficiently large.
This apparent inconsistency seems to have led many to the conclusion that
conventional perturbative evolution breaks down at small $x$.

In fact the dilemma is resolved by the observation that
the physically relevant quantity, namely the splitting function
$P^{gg}(x)$, is instead given at leading order in the expansion
\smxexpsf\ by the series
\eqn\ggspf
{P^{gg}(x;t)={\alpha_s(t)\over x}\sum_{n=1}^\infty (A_n^n)^{gg}
{\left(\alpha_s(t)\ln{1\over x}\right)^{n-1}\over (n-1)!}
\equiv\frac{C_A}{\pi}\frac{\alpha_s(t)}{x}
B\big(\lambda_s(t)\ln\smallfrac{1}{x}\big),}
where now $\lambda_s(t)\equiv\lambda\big(\alpha_s(t)\big)$. This series
converges uniformly on any finite intervals of $x$ and $t$ which
exclude $x=0$. To prove this, we
write the two series \ggls\ and \ggspf\ as
\eqn\twoseries
{A(v)=\sum_{n=1}^\infty a_n v^n, \qquad
B(u) =\sum_{n=1}^\infty \frac{a_n}{(n-1)!} u^{n-1},}
where the coefficients $a_n\equiv 4\ln 2(A_n^n)^{gg}
(4\ln 2\, C_A/\pi)^{-n}$.
The series for $A(v)$ then has radius of convergence one, from which
it follows trivially that the radius of convergence of $B(u)$ is
infinite.

The defining relation \mel\ may now be rewritten in the form
\eqn\borel
{A(v)=\int_0^\infty \! du\, e^{-u/v}\, B(u):}
$B$ is the Borel transform of $A$. The series $A(v)$ is not Borel
summable for $\Re v\inv <1$ (i.e. $\Re N <\lambda$) because the
integral no longer converges at the upper limit (corresponding to the
lower limit of the integral over $x$ in \mel): the real coefficients $a_n$
are all positive and decrease only very slowly as $n\to\infty$, so
$B(u)\sim e^u$ as $|u|\to\infty$.\foot{For timelike anomalous
dimensions, relevant for the small $x$ evolution of fragmentation
functions, the corresponding series have alternating coefficients, and
thus converge throughout the physical region. The same is true of
$\phi^3_6$ theory.} It follows that the only reason for the bad
behaviour of the series \ggls\ is that when transforming to Mellin
space one attempts to integrate all the way down to $x=0$, and this is
not possible for singlet distributions because the total number of partons
diverges there. If instead the parton distributions are evolved using the
Altarelli-Parisi equations \evolAP, the splitting functions are only
required over the physically accessible region $x>x_{\rm min}$ \AP, and
no convergence problems arise. Indeed for all physical applications it
is sufficient to truncate the series \ggspf\ after a finite number of
terms.\foot{In practice this means that
when solving the evolution equations \evol\ numerically, the
divergence of the series expansion of the anomalous dimensions at
small $N$ can simply be ignored: although adding another term in the
series \ggls\ can change the anomalous dimension a lot at small $N$,
this will be compensated by a corresponding shift in the steepest
descent contour, such that the change in the evolved distribution is actually
very small. We have checked this numerically.}

Similar considerations presumably apply to subleading terms in the
series \smxexpad\ and \smxexpsf. Recently the first nonvanishing term in
the large $N$ expansions of $\gamma^{qg}$ and $\gamma^{qq}$ have been
calculated \CHad: the result may be written in the form
\eqn\qgad
{\gamma_N^{qg}(\alpha)=T_R\frac{\alpha}{3\pi}(1+\tilde A(\lambda/N))+O(\alpha),
\qquad \gamma_N^{qq}(\alpha)
=\frac{C_F T_R}{C_A}\frac{\alpha}{3\pi}\tilde A(\lambda/N)+O(\alpha),}
where $\tilde A(v)$ has the same form as $A(v)$ \twoseries, with
new coefficients $\tilde a_n$, but the same radius of convergence.
It follows immediately that the corresponding expressions for $P^{qg}(x;t)$
and $P^{qq}(x;t)$ are convergent for all nonzero $x$.
It is tempting to conjecture that the same must be true for all
the coefficients in the small $x$ expansion \smxexpsf: all that is
necessary is that the coefficients in the expansion of the
corresponding anomalous dimensions \smxexpad\ have a positive radius of
convergence, i.e. that there is no singularity (or accumulation of
singularities) at $N=\infty$, or equivalently at $\alpha=0$. This is a
conventional assumption in perturbative QCD.

Using the analytic expressions given in \refs{\BFKLad,\CHad} we have
computed the coefficients $a_n$ and $\tilde a_n$ to very high
orders. In practice only the first dozen or so are needed in any
realistic calculation: we give the first thirty six in the table.

We are now finally able to write down evolution equations for parton
distribution in the leading order of the small $x$ expansion of
the splitting functions \smxexpsf. Because only the
$P^{gg}$ and $P^{gq}$ splitting functions are nonvanishing at leading
order, only the Altarelli-Parisi equation
satisfied by the gluon distribution survives; the determination of the
quark distribution, and thus of structure functions, requires the
inclusion of subleading corrections and will be discussed at the end
of the paper. As in \DAS\ the evolution equation may be simplified by
introducing the variables
\eqn\exizeta{\xi\equiv\ln\left(\frac{x_0}{x}\right),
       \qquad\zeta\equiv\ln\left(\frac{t}{t_0}\right)
                       =\ln\left(\frac{\ln Q^2/\Lambda^2}
                                      {\ln Q_0^2/\Lambda^2}\right),}
and defining $G(\xi,\zeta)\equiv xg(x;t)$. Using the splitting function
\ggspf\ in the evolution equation \evolAP\ then gives
\eqn\gapeq
{\eqalign{{\partial \over \partial \zeta} G(\xi,\zeta)
&= {4C_A\over \beta_0}\sum_{n=1}^\infty
a_n\frac{\lambda_s(\zeta)^{n-1}}{(n-1)!}
\int_{-\xi_0}^\xi \!d\xi^\prime\,(\xi-\xi^\prime)^{n-1}
G(\xi^\prime,\zeta)\cr
&= {\gamma^2}\sum_{n=0}^\infty a_{n+1}\lambda_s(\zeta)^{n}
\int_{-\xi_0}^\xi \!d\xi_1\,\int_{-\xi_0}^{\xi_1} \!d\xi_2\,\dots
\int_{-\xi_0}^{\xi_n} \!d\xi^\prime\,
  G(\xi^\prime,\zeta).\cr}}
Here $\gamma\equiv\sqrt{4C_A/\beta_0}$ (as in \DAS): we have used the one
loop form of $\alpha_s(t)={4\pi\over \beta_0 t}$,
$\beta_0=11-{2\over 3} n_f$, as is appropriate for a leading order
calculation.
Since we retain only singular contributions to the splitting
functions the lower limit $\xi_0=\ln{1\over x_0}$ in the integrations
on the \rhs\ can be consistently set to zero.

Note that if the large $\xi$ (i.e. small $x$) limit is approached by letting
$\xi$ grow as $\zeta^k$ with any $k>1$, but more slowly than
$e^\zeta$, corrections to the leading order splitting function are
exponentially suppressed. If, however, we let $\xi$ grow faster than
$e^\zeta$ (i.e. $1/x$ grow faster than a power of $Q^2$), then higher
orders in the leading singularity expansion of the splitting function
will eventually become significant, growing as
$e^{\xi\lambda_s(\zeta)}$.

Differentiating both sides with respect to $\xi$,
the gluon evolution equation eq.~\gapeq\ can be cast in the form of
a wave equation as in ref.\DAS:
\eqn\aoweq
{\eqalign{{\partial^2\over\partial\xi\partial\zeta} G(\xi,\zeta)&=
\gamma^2 G(\xi,\zeta)+ \gamma^2 \sum_{n=1}^\infty
a_{n+1}\lambda_s(\zeta)^n
\int_{0}^\xi \!d\xi_1\,\int_{0}^{\xi_1} \!d\xi_2\,\dots
\int_{0}^{\xi_{n-1}} \!d\xi^\prime\,
 G(\xi^\prime,\zeta)\cr
&=\gamma^2 G(\xi,\zeta)+\gamma^2 \sum_{n=1}^\infty
a_{n+1}\frac{\lambda_s(\zeta)^n}{(n-1)!}
\int_{0}^\xi \!d\xi^\prime\,(\xi-\xi^\prime)^{n-1}G(\xi^\prime,\zeta).\cr}}
If only the first leading
singularity is retained, then eq.~\aoweq\ is simply the wave equation
discussed in ref.~\DAS\ (with $\delta=0$; subleading corrections have
not yet been included), and leads directly to a double scaling
behaviour of the gluon distribution when the boundary conditions are
sufficiently soft. When the higher singularities are also included
on the \rhs, then the generic features of the propagation in the
$\xi$--$\zeta$ plane due to the wave-like nature of the
evolution are preserved. In particular, the propagation is still linear
and causal, and far from the boundaries independent of the detailed
form the boundary condition.

For any reasonable values of $\xi$ it is clear that only the first
few terms in the expansion of eq.~\ggspf\ will contribute (notice that in fact
$a_2=a_3=a_5=0$), and then the asymptotic
form of the solution will be essentially unchanged (i.e., it will display
double asymptotic scaling). The easiest way of seeing this explicitly
is to recall that the leading order solution in which only the first
term on the \rhs\ is retained is just given
by a Bessel function $G_0(\xi,\zeta)\sim I_0(2\gamma\sigma)$,
where, as in \DAS, we find it useful to define the scaling variables
$\sigma\equiv\sqrt{\xi\zeta}$, $\rho\equiv\sqrt{\xi/\zeta}$. The
asymptotic behaviour $I_0(z)\sim z^{-1/2}e^z$ as $z\to\infty$ then
gives rise to double asymptotic scaling. If this is substituted back into
the new terms on the \rhs, to generate the first term in an iterative
solution, we may use the fact that $z^{n}I_{n-1}(z)=
\smallfrac{d}{dz}\big(z^nI_n(z)\big)$ to write \aoweq\ in the
approximate form
\eqn\aoweqnapprox{
{\partial^2\over\partial\xi\partial\zeta} G(\xi,\zeta)\simeq
\gamma^2 G(\xi,\zeta)+\gamma^2 \sum_{n=1}^\infty
a_{n+1}\bigg(\frac{\rho\lambda_s(\zeta)}{\gamma}\bigg)^n I_n(2\gamma\sigma).
}
Since $\rho^n I_n(2\gamma\sigma)$ is bounded above by $\xi^n
I_0(2\gamma\sigma)$ it follows that double scaling will always set in
asymptotically provided the series $\sum_{n=1}^\infty a_{n+1}
\big(\smallfrac{\xi\lambda_s(\zeta)}{\gamma}\big)^n$ converges
uniformly, that is provided $\xi<\frac{\gamma}{\lambda_s(\zeta)}$.

However for very large $\xi$ the splitting function will
eventually be dominated by the higher orders of the
series eq.~\ggspf. This is the limit which is often approached by means
of the BFKL equation \Lip, which resums all leading logs of $1\over x$,
i.e. all terms with $p=r$, $q=0$ in \lls, but with the $Q^2$
dependence of $\alpha_s$ suppressed. Since at leading order
eq.\aoweq\ already includes all the leading
twist information contained in the BFKL equation (but with all
higher twist and infrared behaviour factored out), it should reproduce the
same power-like rise in the limit $\xi\to\infty$,
provided that we freeze the coupling. We can both check this, and
understand how the behaviour of the gluon evolution changes when
the coupling runs, by
taking advantage of the fact that when the series eq.~\ggspf\ is
dominated by its higher order terms, the evolution equation
eq.~\aoweq\ takes a simple closed form.

Because the series \ggls,\twoseries\ has unit radius of convergence
it follows that $\neath{\lim}{n\to\infty}
\frac{a_{n+1}}{a_n}\to 1$.\foot{It can be
seen from the table that this asymptotic form begins to set in only for
$n\sim 20$ however.} But then setting $a_{n+1}\approx a_n$ in the sum
in eq.~\aoweq, shifting the summation index by one unit, and using
eq.~\gapeq, we have
\eqn\recur
{\gamma^2\sum_{n=1}^\infty
a_{n+1}\lambda_s(\zeta)^n
\int_{0}^\xi \!d\xi_1\,\int_{0}^{\xi_1} \!d\xi_2\,\dots
\int_{0}^{\xi_{n-1}} \!d\xi^\prime\,
 G(\xi^\prime,\zeta)=\lambda_s(\zeta) {\partial G\over \partial \zeta}.}
Hence, asymptotically as $\xi\to\infty$ the evolution equation \aoweq\
becomes simply
\eqn\damwe
{{\partial^2\over\partial\xi\partial\zeta} G(\xi,\zeta)
- \lambda_s(\zeta){\partial G\over \partial\zeta}=
\gamma^2 G(\xi,\zeta),}
and asymptotically the summation of all leading singularities
leads to a damping term in the wave equation.\foot{As was anticipated,
but not proven, in ref.~\DAS.}

\nref\MRSB{J.~Kwiecinski, A.D.~Martin, W.J.~Stirling and R.G.~Roberts,
               \PR\vyp{D42}{1990}{3645}.}
\nfig\figregval{Regions of validity in the $\xi$--$\zeta$ plane of
various asymptotic behaviours: a) from the damped wave
equation \damwe\ with fixed coupling, b) the same with running
coupling, and c) from the equation \damweeps\ which takes into account the
smallness of the coefficients $a_n$. The region marked `S' denotes the
double scaling region, while `L' denotes the Lipatov (power rise)
region, and `R' the recombination region.}
It is now trivial to recover the singular `Lipatov pomeron'
behaviour: if the coupling is frozen, then
$\lambda_s$ is just a constant, $\lambda$, and the solution to eq.~\damwe\
is given in terms of the (double scaling)
solution $G_0(\xi,\zeta)$ to the original wave equation \DAS\
\eqn\lipsol
{G(\xi,\zeta)= e^{\lambda\xi} G_0(\xi,\zeta)=x^{-\lambda}G_0(\xi,\zeta).}
The solution with fixed coupling thus displays a strong power-like
growth \Lip\ in the region $\xi\gg \frac{\gamma^2}{\lambda^2}\zeta$,
while below this line we have the usual scaling behaviour
(\figregval a).
This situation is very similar to that obtained by solving the original
$\lambda=0$ equation with a hard boundary condition
$G(\xi,0)\sim e^{\lambda\xi}$ \refs{\MRSB,\DAS,\EKL}; however now the
hard behaviour is generated perturbatively,
by the singularity at $N=\lambda$ in the anomalous dimension
eq.~\ggls, rather than imposed by hand.

Even though \damwe\ may be solved exactly for fixed coupling, it will be
useful for the sequel to derive the asymptotic
behaviour of the solution \lipsol, using the usual
technique of Laplace transformation and steepest descent.
Defining ${\cal G}(s,\zeta)\equiv\int_0^\infty\! d\xi
\, e^{-s\xi}G(\xi,\zeta)$, \damwe\ may be written as
\eqn\damlap
{\frac{\partial}{\partial\zeta}{\cal G}(s,\zeta)
=\frac{\gamma^2}{s-\lambda} {\cal G}(s,\zeta)}
which with $\lambda$ fixed integrates to
\eqn\damsollap
{\ln {\cal G}(s,\zeta)=\ln {\cal G}(s,0) + \frac{\gamma^2}{s-\lambda}\zeta.}
Estimating the Bromwich integral by steepest descent,
the saddle point is located at $s_0=\lambda+\gamma/\rho$, and thus for
a soft boundary condition $G(\xi,0)\sim 1$ we find that as $\xi\to\infty$
\eqn\damasymp
{G(\xi,\zeta)\sim \frac{\gamma/\rho}{\lambda+\gamma/\rho}
\frac{1}{\sqrt\sigma}e^{2\gamma\sigma+\lambda\xi}.}
For a hard boundary condition $G(\xi,0)
\sim x^{-\lambda}$ the first factor would be absent: when leading
singularities are included, the form of the boundary condition is only
important very close to the boundary (unless of course the boundary
condition were even harder than $x^{-\lambda}$). This feature persists
in solutions to the full equation \aoweq.

Of course there is really no justification at all for freezing the
coupling: we must solve eq.~\damwe\ with
running coupling $\lambda(\zeta)=\lambda_0 e^{-\zeta}$. Since an
exact solution is no longer available, we work with the
with the Laplace transformed equation \damlap, which now integrates to
\eqn\damsollaprun
{\ln {\cal G}(s,\zeta)=\ln {\cal G}(s,0) + \frac{\gamma^2}{s}
\Big(\zeta +\ln\bigg(\frac{s-\lambda(\zeta)}{s-\lambda_0}\bigg)\Big),}
and we find a branch cut from $\lambda(\zeta)$ to $\lambda_0$.
The saddle point condition leads to a transcendental equation, so
we now treat the large and small $\xi$ limits separately. When
$s$ is large, we can ignore the cut, expand out the logarithm, and
find a saddle point at $s_0 = \frac{\gamma}{\rho} -
\frac{\lambda_0-\lambda}{\zeta}+\cdots$, which gives the asymptotic
behaviour
\eqn\damasymprundas
{G(\xi,\zeta)\sim \frac{1}{\sqrt\sigma}
e^{2\gamma\sigma+(\lambda_0-\lambda(\zeta)\rho^2},}
i.e. double scaling up to a small correction.
This holds throughout the region where the correction is small,
i.e. for $\xi\ll \frac{\gamma^2}{(\lambda_0-\lambda)^2} \zeta^3$. The behaviour
for large $\xi$ is found by dominating the Bromwich integral by the
branch point singularity at $s=\lambda_0$: the saddle point is then at
$s_0=\lambda_0+\frac{\gamma^2}{\lambda_0\xi}+\cdots$, which gives
\eqn\damasymprunlip
{G(\xi,\zeta)\sim \frac{1}{\xi}
\big(\xi(\lambda_0-\lambda(\zeta))\big)^{\gamma^2/\lambda_0}
e^{\xi\lambda_0 + \gamma^2\zeta/\lambda_0}.}
Again this behaviour is valid whenever the corrections to it are small,
which means that $\xi\gg\frac{\gamma^2}{\lambda_0^2}e^\zeta$.
Thus the running of the coupling leaves the exponent of the power-like
growth unchanged (at $\lambda_0$), but severely limits the region in
which this is the
dominant behaviour (see \figregval b). The region in which the scaling
behaviour \damasymprundas\ obtains is correspondingly increased.

In practice the damped wave equation \damwe\ is still a rather poor
approximation to the full equation \aoweq, since, as may be seen from
the table, the ratio $\frac{a_n}{a_1}\ll 1$ for all $n>1$. A better
approximation may thus be obtained by setting $a_n = \epsilon a_1$;
this gives the rather more complicated equation
\eqn\damweeps
{\bigg(\frac{\partial}{\partial\xi} - \lambda\bigg)
\bigg({\partial^2\over\partial\xi\partial\zeta} -\gamma^2\bigg)G(\xi,\zeta)
=\epsilon\gamma^2\lambda G(\xi,\zeta),}
which however on Laplace transformation becomes simply
\eqn\damlapeps
{\frac{\partial}{\partial\zeta}{\cal G}(s,\zeta)
=\frac{\gamma^2}{s}\bigg(1+\frac{\epsilon\lambda}{s-\lambda}\bigg)
{\cal G}(s,\zeta).}
The asymptotic behaviour of the solution may be found just as
above: for running coupling \damlapeps\ again integrates to \damsollaprun,
but with the logarithm now suppressed by a factor of epsilon, which
means that the double scaling solution \damasymprundas\ holds
throughout the even larger region $\xi\ll \frac{\gamma^2}{\epsilon^2
(\lambda_0-\lambda)^2} \zeta^3$, the subleading term in the exponent
being suppressed by $\epsilon$. The power-like growth
\damasymprunlip\ is now confined to the region $\xi\gg
\frac{\epsilon\gamma^2}{\lambda_0(\lambda_0-\lambda)}e^{\zeta/\epsilon}$
(see \figregval c).
Since $\epsilon\lsim 0.1$, this means that unless $\zeta$ is
very small, the Lipatov growth only sets in when $\xi$ is exceedingly
large.\foot{For example if $\zeta\sim 0.4$ (corresponding to $Q^2$
around $10 \GeV^2$), we would need $\xi$ to be of order $100$.}

\nfig\figcon{Contour plots of the logarithms of the solutions of
the small-$x$ evolution
equations in the $\xi$--$\zeta$ plane: a) the gluon distribution
$G$, b) the structure function $F_2$ with $Q_0 = 1~\GeV$, and c) the
same but with $Q_0 = 2~\GeV$. The values of the other parameters are
explained in the text. The contours are all equally spaced: the
interval between adjacent contours is $\Delta\ln G=0.85$, $\Delta\ln
 F_2 = 0.62$,~$0.58$ respectively. A scatter plot of the HERA
data\refs{\ZEUSc,\Honec} is superimposed for reference.}
These results are confirmed by a numerical analysis of the full
evolution equation Eq.~\aoweq. The equation was solved by using the
integrated form
\eqn\Goursat
{\eqalign{G(\xi,\zeta)=G_0(\xi,\zeta) &+
\int_0^\xi\!d\xi'\,\int_0^\zeta\!d\zeta'\,
I_0\big(2\gamma\sqrt{(\xi-\xi')(\zeta-\zeta')}\big)\cr
&\times\,\gamma^2\sum_{n=1}^\infty
a_{n+1}\frac{\lambda_s(\zeta')^n}{(n-1)!}
\int_{0}^{\xi'} \!d\xi''\,(\xi'-\xi'')^{n-1}G(\xi'',\zeta'),\cr}}
where $G_0(\xi,\zeta)$ is the solution of the leading order equation,
as given in \DAS. In practice the series on the \rhs\ converges to an
accuracy of less than one per cent after only a dozen or so terms.
Eqn.\Goursat\ may be solved by the standard iterative procedure:
convergence is achieved in practice after only a few iterations.
For definiteness we chose $n_f=4$, $\Lambda=120 \MeV$ (so that
$\alpha_s(M_z)=0.116$), soft boundary conditions $G(\xi,0)=G(0,\zeta)=
{\rm const.}$ and $x_0=0.1$, $Q_0=1\GeV$. The
result is shown in the contour plot \figcon a, to be compared with the
leading order plot fig.~3a of ref.\DAS. Even though the range of $\xi$
has been increased, there are no significant scaling violations except
very close to the left hand axis.

\nref\GLR{ L.V.~Gribov, E.M.~Levin
and M.G.~Ryskin, \PRep\vyp{100}{1983}{1}.}
In summary, our analysis of the
small $x$ all-order gluon evolution equation eq.~\aoweq\
shows that the leading order double scaling behaviour \DAS\ holds
asymptotically throughout most of the $\xi$--$\zeta$ plane, the
power-like singular behavior eq.~\lipsol\ being confined to a
very small wedge close to the boundary $\zeta=0$. This is to be
contrasted with the situation for fixed coupling \Lip, or
the phenomenological approach in which a singular
behaviour is input to the leading order evolution equation
\refs{\MRSB,\DAS,\EKL}: in both cases the
power like singularity is preserved by the evolution and propagates into
a large region $ \xi \lsim \smallfrac{\gamma^2}{\lambda^2}\zeta$. The
reason that the region in which the power-like growth develops is in
reality so small and narrow is basically twofold: the
running of the coupling means that the growth near the boundary
is rapidly damped away by the exponential fall of $\alpha_s$ as
$\zeta$ increases, and the size of the growth is in any case severely
limited by the smallness of the coefficients $a_n$ of the leading
singularities. It thus seems extremely unlikely that the
`hard pomeron' will be seen in structure functions, since at any
reasonable value of $Q^2$, the power-like rise only sets in at
extremely small values of $x$, $\ln\frac{1}{x}\gg
\big(\frac{\alpha_s(t_0)}{\alpha_s(t)}\big)^{1/\epsilon}$,
$1/\epsilon \gsim 20$, and in any case such small values of $x$ are
well inside the recombination region \refs{\GLR,\Catrev} $\ln\frac{1}{x}\gg
\ln\frac{1}{x_r}\big(\frac{1}{\alpha_s(t)}\big)^2$, $x_r\lsim 10^{-4}$
where perturbation theory has already broken down, and the rise is
presumably damped
by the nonperturbative effects necessary to restore unitarity \DGPTWZ.

Having found the gluon distribution, we may now discuss the determination
of the quark distribution and the structure function $F_2(x;t)$, given
in the parton scheme by
\eqn\eFtwo{\eqalign{F_2^p(x;t)
&=\smallfrac{5}{18}Q(x;t)+F_2^{NS}(x;t)\cr
Q(x;t)&\equiv x\sum_{i=1}^{n_f}\big(q_i(x;t)+\bar{q}_i(x;t)\big).\cr}}
In the small $x$ region, the nonsinglet contribution $F_2^{NS}(x;t)$
may be neglected. Since $\gamma^{qg}$ and $\gamma^{qq}$ begin at
order $\alpha$ (see \qgad) to calculate $Q$ from $G$ we must go to
next-to-leading order in $\alpha$
in the expansion eq.~\smxexpad. Writing this as
$\gamma^{ij}(\alpha)=\sum_{a=0}^\infty \alpha^a\gamma^{ij}_a$, we can
diagonalize $\gamma^{ij}(\alpha)$ order by order in $\alpha$. To
first order the eigenvalues are
\eqn\eval
{\eqalign{\lambda^+(\alpha)&= \gamma_0^{gg}
+\alpha
\big(\gamma_1^{gg}+\frac{\gamma_0^{gq}}{\gamma_0^{gg}}\gamma_1^{qg}\big)
+O(\alpha^2)\cr
\lambda^-(\alpha)&=\alpha\big(\gamma_1^{qq}-
\frac{\gamma_0^{gq}}{\gamma_0^{gg}}\gamma_1^{qg}\big)+O(\alpha^2).\cr}}
The corresponding eigenvectors are given by
\eqn\evec
{Q^+ = \alpha \frac{\gamma_1^{qg}}{\gamma_0^{gg}}G^+
+O(\alpha^2)\qquad Q^- = - \frac{\gamma_0^{gg}}{\gamma_0^{gq}}G^-
+O(\alpha).}
The eigenvectors are not orthogonal, because $\gamma^{ij}$ is not
symmetric.
At leading order the larger eigenvalue $\lambda^+$ leads to the
wave equation \aoweq\ for $G^+$, while the small eigenvalue
vanishes, so $G^-$ and $Q^-$ do not evolve. At order
$\alpha$, \qgad\ gives
$\lambda^-(\alpha)=-\frac{\alpha}{3\pi}\frac{C_F T_R}{C_A}$,
so $G^-_N(t)=G^-_N(0)e^{-\delta'\zeta}$,
$Q^-_N(t)=Q^-_N(0)e^{-\delta'\zeta}$, where
$\delta'=\frac{16n_f}{27}\beta_0\inv$. The full correction to
$\lambda^+$ is as yet unknown, since only the first two terms of
$\gamma_1^{gg}$ have been computed. Retaining only the first (one
loop) term the evolution equation \aoweq\ is supplemented by a damping term
proportional to $\delta \equiv (11+\frac{2n_f}{27})/\beta_0$ as in
\DAS. The solutions $G^+(\zeta)$ and $Q^+(\zeta)$
then acquire an extra factor of $e^{-\delta\zeta}$.

Using the eigenvalues \eval\ and eigenvectors \evec, and the expansion
\qgad\ for $\gamma_1^{qg}$, it is now not
difficult to derive the quark
evolution equation: taking as boundary conditions $G(\xi,0)\equiv
G_0(\xi)$, $Q(\xi,0)\equiv Q_0(\xi)$,
\eqn\qg
{\eqalign{\frac{\partial}{\partial\zeta}Q(\xi,\zeta)
=-\delta' Q_0(\xi)&e^{-\delta'\zeta}
+\frac{n_f}{9}\gamma^2 e^{-\delta\zeta}\Big[\bar G(\xi,\zeta)\cr
&+\sum_{n=1}^\infty \tilde a_n \lambda_s(\zeta)^n\int_0^\xi\! d\xi'\,
\frac{(\xi-\xi')^{n-1}}{(n-1)!} \bar G(\xi',\zeta)\Big],\cr}}
where the gluon distribution $\bar G(\xi,\zeta)$ is a solution of the
wave equation \aoweq\ with boundary condition $\bar G(\xi,0)=G_0(\xi)+
\smallfrac{4}{9}Q_0(\xi)$. This equation may then be readily
integrated to give the quark distribution and thus $F_2^p$. Although
we have derived this equation in
the parton scheme, it is actually scheme independent: changes in the
renormalization scale only affect terms which are of higher order in
$\alpha_s$.\foot{This may be proven explicitly; although in a
different scheme (such as $\overline{\rm MS}$) the coefficients
$\tilde a_n$ will be
modified, the difference is made up by the new $O(\alpha_s)$
corrections to the coefficient functions \CHad.} In this sense the
expression \qg\ is still leading order, and should thus be treated on
the same footing as \aoweq; scheme dependent subleading corrections
first come in through higher loop singularities in $\gamma_1^{gg}$.

We may now discuss the asymptotic behaviour of the quark distribution.
Since throughout most of the $\xi$--$\zeta$ plane $\bar G(\xi,\zeta)\simeq
{\cal N}I_0(2\gamma\sigma)$, we may use the same trick as we did to derive
\aoweqnapprox\ to write \qg\ as (setting $\delta$ and $\delta'$ to
zero for simplicity)
\eqn\qgapprox
{\frac{\partial}{\partial\zeta}Q(\xi,\zeta)\simeq\frac{n_f}{9}\gamma^2
{\cal N}\sum_{n=0}^\infty \tilde a_n
\bigg(\frac{\rho\lambda_s(\zeta)}{\gamma}\bigg)^n I_n(2\gamma\sigma).}
It follows that when $\xi\ll\frac{\gamma^2}{\lambda_0^2}\zeta
e^{2\zeta}$, $Q(\xi,\zeta)$ scales just as observed experimentally
\refs{\DAS,\Test}. For
$\xi\gg\frac{\gamma^2}{\lambda_0^2}\zeta e^{2\zeta}$, if we set
$\tilde a_n = 1$ for all $n$, \qgapprox\ may be simplified yet further
by using the relation $\sum_{-\infty}^{\infty}t^n I_n(z) =
\exp\big(\half z(t+t\inv)\big)$, to give
\eqn\qpower
{Q(\xi,\zeta)\sim {\cal N}\zeta
e^{\xi\lambda_0+\gamma^2\zeta/\lambda_0}.}
This result may be readily confirmed using Laplace transforms with
respect to $\xi$.

It follows that even when the gluon scales, $F_2$ can still grow rapidly
sufficiently close to the $\zeta = 0$ boundary because of higher order
singularities in the anomalous dimension $\gamma^{qg}_N$. Indeed these
singularities turn out to be much more important than those of
$\gamma^{gg}_N$, despite being suppressed by an additional power of
$\alpha$, essentially because the coefficients $\tilde a_n\gg a_n$.
Note that in the BFKL equation \Lip\ quark loops are suppressed, so the
contribution of these singularities is not included. In fact even if it were
possible to detect the tail of a power-like rise in the gluon
distribution, in the structure function this would be masked by a
similar rise generated by the subleading singularities in the coupling
of the gluons to quarks. This effect may be confirmed
numerically by evaluating \qg, using as input the gluon distribution
displayed in \figcon a, and taking as boundary condition
$Q_0\simeq\half G_0$. The resulting contour plot of $\ln F_2$ is
displayed in \figcon b. Indeed $F_2$ differs very
little from the double scaling result except in the wedge close to the
left hand boundary, where instead the behaviour \qpower\ is found.

The growth \qpower\ should not be taken to literally, however, since
for any reasonable value of $x$ it only occurs for values of $Q^2$
very close to the starting scale $Q_0^2$, at which nonperturbative
effects are also relatively important. Furthermore, in this region
higher twist recombination effects are expected to become
significant \GLR. More interestingly, the
higher order terms in \qg\ increase the relative
normalization of $Q$ with respect to $G$ at small $x$ for a given
starting scale $Q_0$ (by around a factor of two). Since when $Q_0 =
1~\GeV$ the leading order result has approximately the correct
normalization, this suggests that at all orders the starting scale
$Q_0$ at which the soft pomeron boundary condition is imposed should
be raised to compensate. In fact, a starting scale of around
$2~\GeV$ turns out to be required; the
resulting contour plot of $\ln F_2$ is shown in \figcon c. Since
$\lambda_0$ is a little lower, the scaling behaviour is also
improved, the power-like growth being now confined to the top
left-hand corner of the plot.

To make the comparison with the double scaling results of ref.~\Test\
more quantitative, we may compute the slope of the linear rise of
(in the notation of \refs{\DAS,\Test}) $\ln R_F'F_2$ in the HERA region.
In leading order this is simply given by $2\gamma = 2.4$, which is
reduced by around $20\%$ by two loop corrections \Mont, whereas the
HERA data give $2.4\pm 0.2$. The corresponding slopes for the three
distributions plotted in \figcon\ are $2.5$, $2.9$ and $2.6$
respectively. Clearly a detailed fit to the data would thus require the
inclusion of all of these competing effects.

In conclusion, the leading twist evolution equation which sums all
leading logarithms at small $x$, but is also fully consistent with the
renormalization group, interpolating smoothly between small $x$ and
large $x$, is just the Altarelli-Parisi equation \evolAP,
with the splitting function expanded in powers of $\alpha_s(t)$, but
retaining terms to all orders in
$\alpha_s(t)\ln\frac{1}{x}$. Perturbation theory at small $x$ is then
placed on the same footing as perturbation theory at large $x$: there
are no instabilities or problems of convergence.
In fact deep inelastic scattering at small $x$ and large $Q^2$ is a
much cleaner perturbative environment than more traditional deep
inelastic scattering at large
$x$, since the non-perturbative component of the structure function is
relatively insignificant \refs{\DAS,\Mont}. Taken altogether, the
analysis presented here
explains why the simple double asymptotic scaling
picture presented in \DAS, though only at leading (and subleading
\Mont) order in $\alpha_s$, accounts so well \Test\ for the rise in
$F_2^p$ seen at HERA \refs{\ZEUSc,\Honec}.
Perturbation theory breaks down only when higher twist effects
become necessary to restore unitarity.

Although they have little effect on the qualitative form of $F_2$ for any
reasonable values of $Q^2$, the higher order logs in $P^{qg}$ do make
a significant impact on the relative normalization of the
quark and gluon distributions. This means that attempts to extract
the gluon distribution from $F_2$ using
the standard two-loop evolution equation should be
treated with caution: an all-order calculation could give a gluon
distribution which is smaller by as much as a factor of two. The
summation of subleading logarithms will become
increasingly important as the quality and range of structure function
data at small $x$ improves further.

\medskip
{\bf Acknowledgement:} We are very grateful to S.~Catani for several useful
discussions on this subject.
\vfill
\eject
\listrefs

\topinsert\hfil
\vbox{
      \baselineskip\footskip
     {\tabskip=0pt \offinterlineskip
      \def\tablerule{\noalign{\hrule}}
      \halign to 300pt{\strut#&\vrule#\tabskip=1em plus2em
                   &\hfil#\hfil&\vrule#
                   &#\hfil&\vrule#
                   &\hfil#&\vrule#\tabskip=0pt\cr\tablerule
             &&\omit\hidewidth $n$\hidewidth
             &&\omit\hidewidth $a_n$\hidewidth
             &&\omit\hidewidth $\tilde a_n$\hidewidth
              &\cr\tablerule
&&   1 && 1          && 0.78145981 & \cr 
&&   2 && 0          && 0.29913310 & \cr 
&&   3 && 0          && 0.38806675 & \cr 
&&   4 && 0.11279729 && 0.25256339 & \cr 
&&   5 && 0          && 0.17838311 & \cr 
&&   6 && 0.01265760 && 0.22632345 & \cr 
&&   7 && 0.03816969 && 0.15473331 & \cr 
&&   8 && 0.00160119 && 0.13891462 & \cr 
&&   9 && 0.01142194 && 0.15744938 & \cr 
&&  10 && 0.01742873 && 0.11602071 & \cr 
&&  11 && 0.00260717 && 0.11532868 & \cr 
&&  12 && 0.00888439 && 0.11997040 & \cr 
&&  13 && 0.00942680 && 0.09568020 & \cr 
&&  14 && 0.00300214 && 0.09846280 & \cr 
&&  15 && 0.00670932 && 0.09707906 & \cr 
&&  16 && 0.00581292 && 0.08291446 & \cr 
&&  17 && 0.00301620 && 0.08559381 & \cr 
&&  18 && 0.00507784 && 0.08203738 & \cr 
&&  19 && 0.00400435 && 0.07384063 & \cr 
&&  20 && 0.00283184 && 0.07550427 & \cr 
&&  21 && 0.00390386 && 0.07157467 & \cr 
&&  22 && 0.00301599 && 0.06682914 & \cr 
&&  23 && 0.00256389 && 0.06747701 & \cr 
&&  24 && 0.00306948 && 0.06392641 & \cr 
&&  25 && 0.00242542 && 0.06112241 & \cr 
&&  26 && 0.00227736 && 0.06101759 & \cr 
&&  27 && 0.00247526 && 0.05808061 & \cr 
&&  28 && 0.00203856 && 0.05633377 & \cr 
&&  29 && 0.00200551 && 0.05576012 & \cr 
&&  30 && 0.00204715 && 0.05343445 & \cr 
&&  31 && 0.00176238 && 0.05224401 & \cr 
&&  32 && 0.00176276 && 0.05142727 & \cr 
&&  33 && 0.00173286 && 0.04961887 & \cr 
&&  34 && 0.00155085 && 0.04871349 & \cr 
&&  35 && 0.00155308 && 0.04780818 & \cr 
&&  36 && 0.00149656 && 0.04640235 & \cr 
\tablerule}}}
\hfil\bigskip
\centerline{\vbox{\hsize= 200pt \raggedright\noindent\footnotefont
Table: The coefficients $a_n$ and $\tilde a_n$, computed using
formulae in ref.\BFKLad\ and ref.\CHad\ respectively.
}}
\bigskip
\endinsert
\vfill
\eject

\listfigs
\bye